\documentclass[english]{article}
\usepackage[T1]{fontenc}
\usepackage[latin9]{inputenc}
\usepackage{url}
\usepackage{amsmath}
\usepackage{graphicx}

\makeatletter
\usepackage{watermark}

\makeatother

\usepackage{babel}
\begin{document}

\title{A Stable Coin with Pro-rated Rebasement and Price Manipulation Protection }

\author{Jose I. Orlicki - jose@toroidtoken.org}

\date{05/07/2017}
\maketitle
\begin{abstract}
An existing pseudo-commodity and a smart contracts framework allow
the creation of a purely automatic and self-sufficient price-stable
cryptocurrency, without human intervention. This new currency, we
denominated Toroid\footnote{\url{} v0.106 UNPUBLISHED WORK: All rights reserved. This work is
not to be published in any way without permission from the author. } or TRD, can be used more extensively for commerce than pseudo commodity
cryptocurrencies due to its lower volatility. Also, is suitable for
investment, as the tokens in each account multiply, return interest,
when the market grows. Like the controlled fiat money of a central
bank plus the benefits of an inflation-adjusted perpetuity bond. It
eliminates the need of debt as the basic mechanism of the economy,
although allows banking and further lending on top of it. Collateral
in base coin, for example BTC or ETH, can be added to bootstrap your
own Toroid investment or withdrawed after a very small investment
period. So, the Toroids are not created from nothing nor have a limited
monetary base. The minimum investment period can be very small, for
example one day, and you keep the interest but you can return the
Toroids and refund your collateral. That is a one-side only peg to
a deflationary crypto-commodity. The stability is guaranteed by endogenous
measurements of number of transactions and wallet pro-rated rebasement
of balance to reduce volatility of price. Each account has its own
rebasement due to the account creation timestamp. Rebasement control
mechanism is progressive during initial bootstrap period because price
manipulation protection is more severe when the capital involved is
smaller. Rebasement has a quick positive start to incentivize early
adopters that see only big growth in their TRD account during bootstrap
period. Finally, the new rebasement control makes it economically
infeasible for an attacker targeting the coin with manipulated transaction
volume if we set the minimum rebasement greater than profits from
massive currency manipulation. This protection increases its robustness
and effectiveness over time because a coin with bigger capitalization
is more difficult to manipulate.
\end{abstract}
\newpage{}

\section{Introduction}

There is some discussion that we need new cryptocurrencies that can
keep the pace of the economy, that can grow with the economy. Possible
best alternatives are not based on debt. But still need to be dynamic,
as pseudo commodities like Bitcoin have an almost static supply after
some time.
\begin{quotation}
With Ether, the money supply grows faster than BTC, but still slow
enough to have a growth rate that tends towards $0$ (on the order
of $1/t$). Assuming the economy grows faster than a linear rate (which
historically has always been the case), Ether is also a \textquotedbl{}deflationary
currency\textquotedbl{}. \cite{etherdeflation}
\end{quotation}
Current fiat currencies based on debt are facing a huge crisis of
trust due to growth stagnation in the developed countries and citizens
and entrepreneurs demanding more freedom to generate their own financial
assets. Also, many players are demanding a reduction in the financial
friction such as fees, so everyone can invest on small and big scale.
Most important, if big players don't get so unproportioned returns
on investment compared to smaller players that can't afford expensive
sophisticated funds and management services, the later can enter the
game. 
\begin{quotation}
Ultimately, excessive debt resembles a Ponzi scheme. Nations, businesses,
and individuals need to borrow ever-increasing amounts to repay existing
borrowings and maintain economic growth. In the half-century leading
up to 2008, the amount of debt needed to create US\$1 of GDP in the
US increased from US\$1-2 to US\$4-5. This rapid rise is insustainable,
given an aging population, slower growth, and low inflation. \cite{stagnation}
\end{quotation}
Similarly to blood cells that are generated in abundance from bone
marrow stem cells, and similarly to gold or silver-backed paper currencies,
we are proposing that the current robust cryptographic pseudo-commodities
\cite{bitcoin} and the smart contract platform are used as a basis
for pseudo-fiat currencies that will be bootstrapped by depositing
basic coins (will be using ETH in example and first token) to generate
a sub-currency that has enough features to have a more stable price,
to have an non-limited monetary supply and to provide economical unfeasibility
of currency manipulation.

To obtain all the features of currency which are scarcity, fungibility,
divisibility, durability, and transferability, and to obtain the three
main uses of currency which are as a medium of exchange, as a unit
of account and as a store of value, we need our currency to be more
stable than Bitcoin or gold. Who can pay for a coffee in Bitcoin if
the coffee shop owner knows the price can change 10\% in minutes?
Currencies which includes features to stabilize its price are sometimes
called stable coins \cite{stable} or if they use an elastic monetary
base are called Hayek Money \cite{hayek}. Other technical definition
can categorize our stable coin as a smart beta bond \cite{smartbeta},
an intelligent bond with a market exposure $\beta$ lower than 1,
that indicates an investment with lower volatility than the market. 

Some stable coins have been designed to support collateral deposits
to generate bonds in a way to bootstrap the new currency \cite{makerdao}.
In our proposed design, we also use collateral to bootstrap the new
currency but the funds remain frozen only a small period and you have
the liberty to withdraw it but you keep the variable interest rate
you recieved (and can be negative). Also, the stable coin we proposed
is no two-way pegged to an external basket of existing currencies.
Is only one way pegged, price can only be smaller, than the basic
pseudo-commodity that fuels the smart contracts consensus framework.
We also need some mechanism to deflect currency manipulation attacks
or at least to make them economically unfeasible.

\section{Features}

\subsection{General features}

The Toroid cryptocurrency fund can be implemented easily on a cryptocurrency
platform with smart contracts such as RootStock or Ethereum. It will
have a one-way peg to the underlying deflationary token of the platform,
such as BTC (in the RSK platform) or ETH (in the Ethereum platform).
This results on the price of the TRD being smaller than the underlying
pseudo-commodity but its monetary base being elastic. We will call
base coin to the underlying BTC or ETH that fuels the smart contract
platform and serves as collateral for bootstrapping ICOs (Initial
Coin Offerings) and investment on the new more stable currency. Also,
we will call stable coin or TRD to the new sub-currency defined by
the smart contract.

We have an automatic rebasement in place, that is, the amount of TRDs
in circulation is controlled by the smart contract that represents
the fund and all the accounts or wallets. The rebasement is pro-rated
across all wallets. An account with twice as TRD compared to another
account will get twice of the rebasement adjustment measured in TRD
coins. After the initial bootstrap period where is only positive,
the rebasement can be positive or negative. A recurrent time-period,
such as 1 day, 1 hour or 5 minutes is fixed as the TRD monetary base
adjustment. For example, if we choose a 1-day period and 12AM GMT
as the time of the rebasement and the smart contract decided that
the current daily rebasement is 0.1\% positive then all account have
their balance increased on 0.1\% to compensate endogenous metrics
that have been measured on the previous period.

\subsection{One-way peg to underlying deflationary token}

This means that the price of the TRD will always be lower than a fixed
amount of base cryptocurrency (we choose Ethereum platform and ETH
base token for the sake of example). The one-way peg is implemented
as following: 
\begin{enumerate}
\item to open a wallet with 1 TRD you need to deposit a fixed collateral,
for example 1 decibitcoin (1/10 of ETH or 1 dETH );
\item you can buy any amount of extra TRDs by depositing more collateral
with the same rate of 1 dETH;
\item the collateral is held as collateral in the smart contract and associated
with your wallet until you close you decide, if you want, to return
the original TRD amount;
\item to close your account and get back the complete collateral you need
to have at least in your account the amount of TRDs proportional to
the amount of collateral.
\end{enumerate}
The result of this design is that you can buy 1 TRD with 1 dETH or
less, but never the price of 1 TRD will exceed 1 dETH, the fixed amount
of collateral in the configuration. During the bootstrap of the Toroids
there are incentives to deposit collateral and get the TRD interest
added. At some point, during the Toroids will be less expensive on
an exchange and you also get the return of the investment of you buy
them on an exchange. Ideally, during bootstrap and after that during
the first moments of the Toroids TRD owners will have little incentives
to sell their TRD so new users will directly invest collateral to
get their Toroids. If at some point $Price_{TRD}>Price_{ETH}$ the
arbitrageurs will start funding the Toroid Fund to get cheap TRD and
the price of the TRD will drop. If $Price_{TRD}<Price_{ETH}$ investors
will analyze the estimated interest of ETH deposits returned in TRD
to decide if they want to fund with ETH or buy TRD directly in the
open market. If the TRD in circulation are few, then users that want
to refund their ETH will need to buy TRD so the price of TRD will
increase until $Price_{TRD}>Price_{ETH}$ when arbitrageurs will start
funding and increasing the TRD supply.

\subsection{Elastic currency supply with security limits}

On an initial stage the currency price won\textquoteright t be stable
but will grow its supply to incentivize the early adopters. This period
lasts time $T$, for example 3 months. Also, the rebasement is by-design
limited by the cost of an attack, so every attack is economically
infeasible. We can model this by saying that rebasement ratio $R$
consists of three components:

\[
R(t,v,s)=\begin{cases}
R_{I}(t)+\min\{R_{Gas}(t,v,s),\ R_{Vol}(t,v)\} & \text{if }R_{Vol}(t,v)\geq0\\
R_{I}(t)+\max\{-R_{Gas}(t,v,s),\ R_{Vol}(t,v)\} & \text{if }R_{Vol}(t,v)<0
\end{cases}
\]

TRD supply $S$ is rebased accross all wallets, according to

\[
S(t):=S(t-1)*(1+R(t,v))
\]

Volume $v$ is considered only as the number of transactions during
the last period. We do not consider trading volume in TRDs because
with only few big transactions any Sybil attack can be performed by
many users or one user with many wallets. Supply $S$ is the total
supply of TRDs in the system. Time step variable $t$ help us estimated
time from bootstrap and analyze the previous steps during computations.

The initial $R_{I}$ ratio starts for example in 1.10 and slowly converges
to 1.0 as the incentives for early adopters should fade after the
other rebasement components becomes the main component. Example:

\[
R_{I}(t)=1/(t+10)+1
\]

In this case $R_{I}(0)=1.1$ and $R_{I}(90)=1.01$, so after 90 days
the initial incentives per daily period reduces to only 1\% from the
10\% per day at the beginning. This initial rebasement ratio is not
a function of last period\textquoteright s volume of transactions. 

As of $R_{Gas}$, this section of the rebasement places a limit on
the variation based on the cost of the transaction, to make the currency
manipulation economically unfeasible for an attacker that wants to
generate a massive number of fake transactions. As gas, based on the
Ethereum platform, we mean the cost measured in native deflationary
token (ETH) of executing different smart contracts. If we have during
the last period $v$ wallet transactions, considering all the wallets
in the system, and during the previous period there were no transaction
at all\footnote{Remember that this is not transaccion volume, but the number of transactions.
Other factors can also be used.}. This is the worst-case scenario. We assume in this discussion that
the price is gas is 20 gwei and the cost of basic transaction in the
Ethereum platform is 20000 gas. And we consider that each transaction,
a transference of TRDs from one wallet to another, costs a fixed amount
of gas, measured in fractions of underlying base token (in our example
is ETH). Then one basic wallet transference, costs 0.0004 ETH. If
all activity during the period was only a Sybil attack then the cost
of the attack is:

\[
C_{Sybil}=v*0.0004\ \textrm{ETH}\geq v*0.004\ \textrm{TRD}
\]

We used the knowledge that 1 TRD is one-way pegged to 1 deciETH. If
we have $v=10^{4}$, an attack of 10 thousand spurious transactions,
then the cost for the attacker is $C_{Sybil}=4\ \textrm{ETH}\geq40\ \textrm{TRD}$.
So, if we limit the profit from a malicious Sybil rebasement to 40
TRD, in this case, then we make the attack economically infeasible.
Then if total supply $s$ of TRDs:

\[
R_{Gas}\leq40\ \textrm{{TRD}}/s=v*4*10^{-3}\textrm{{TRD}}/s=\frac{v}{s}*0.004\ \textrm{{TRD}}
\]

Then $1+R_{Gas}$ is the ratio allowed to limit Sybil attacks including
an increase in the number of transactions. Similarly, $1-R_{Gas}$if
the number of transactions have been suddenly reduced in one period.
We are assuming that currency supply rebasement is very quickly absorbed
but market demand reflected linearly on price changes. Is a very simple
model but other more sophisticated can be implemented in future versions.
An ideal TRD transaction, hostile or ordinary, will never contribute
more than its gas transaction cost to the total rebasement measured
in ETH.

The main type of attacks to the stabilized cryptocurrency we want
to prevent are \textit{pump and dump}. Happen when simulated or exaggerated
interest on the new coin is showed in the market to inflate its price
and then the inventory of coin is liquidated by the attacker at an
unusually higher price. Because of this type attack also the price
of the coin plummets to very low levels.

\section{Simulated results}

We simulated the behaviour of Toroid price compared to Bitcoin assuming
the market absorbs the rebasement and adjust completely the coin to
a new price. We can see in Figures 1 and 2 the price evolution over
time of a simulated version or Toroid token. We see that the gas limit
protection comes with a drawback, that is, a less powerful control
over the price. Something similar is shown in Figures 3 and 4 where
the token supply is shown, and including the gas limit makes the supply
of the simulated Toroid grow slowly but still exponentially compared
to base Bitcoin.

\begin{figure}
\includegraphics[scale=0.88]{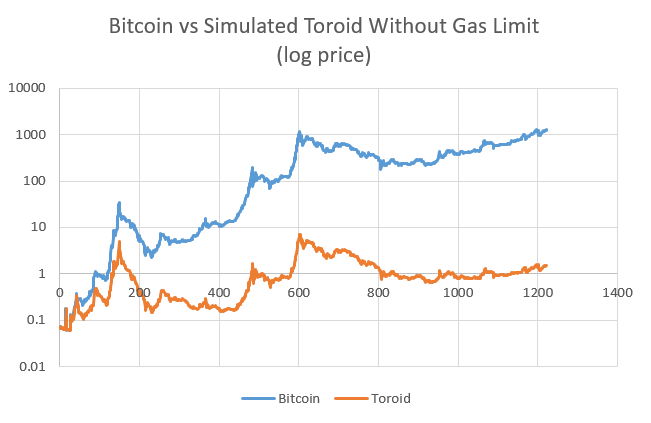}

\caption{Price of Bitcoin versus Price of Simulated Toroid, only with smart
rebasement, from 2010-08-18 to 2017-04-25. }
\end{figure}

\begin{figure}
\includegraphics[scale=0.88]{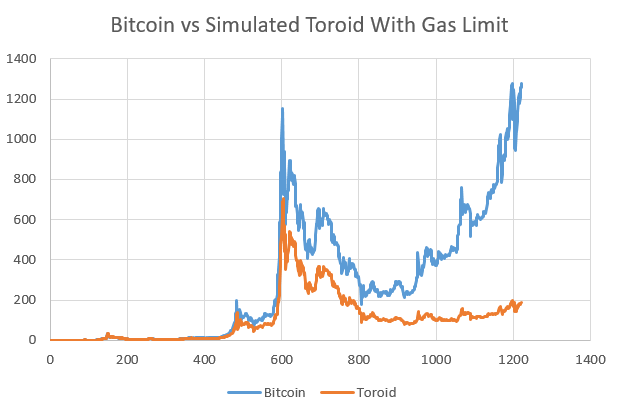}

\caption{Price of Bitcoin versus Price of Simulated Toroid, smart rebasement
and gas limit protection, from 2010-08-18 to 2017-04-25. We assumed
an initial supply of 10000 TRD and gas cost per transaction equivalent
to 0.1 TRD.}
\end{figure}

\begin{figure}
\includegraphics[scale=0.88]{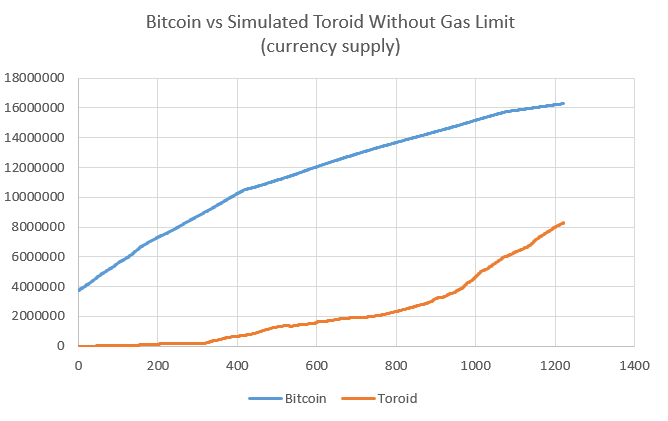}

\caption{Supply of Bitcoin versus Supply of Simulated Toroid, only smart rebasement,
from 2010-08-18 to 2017-04-25. We assumed an initial supply of 10000
TRD.}
\end{figure}

\begin{figure}
\includegraphics[scale=0.88]{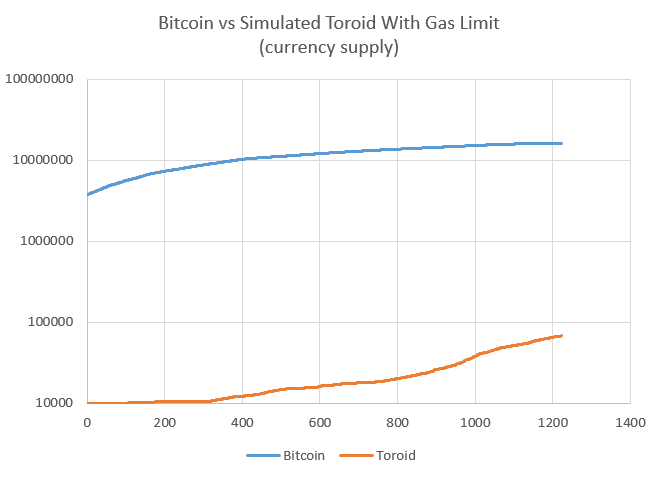}

\caption{Supply of Bitcoin versus Supply of Simulated Toroid, smart rebasement
and gas limit protection, from 2010-08-18 to 2017-04-25. We assumed
an initial supply of 10000 TRD and gas cost per transaction equivalent
to 0.1 TRD.}

\end{figure}

\section{Conclusions}

We proposed a system of stabilized cryptocurrencies based on a Smart
contract platform that allow us to use the currency fueling the Smart
contracts as collateral for the generation of the new coin interest.
This system is not an artificial Ponzi scheme because the interest
generated, based on endogenous metrics, can also be negative, the
holding period can be is very small, and the stabilization of the
new currency is adding value to the new currency and improving its
conditions to be used as means of exchange for goods and services. 

\bibliographystyle{plain}
\nocite{*}
\bibliography{WhitePaper}

\end{document}